\documentclass[12pt]{article}

\usepackage{graphicx}
\usepackage{fullpage}
\usepackage{bbm, amssymb}
\usepackage{amsmath}
\usepackage{eurosym}
\usepackage{verbatim}
\usepackage{natbib}

\numberwithin{equation}{section} 

\newcommand{\DeltaVup}{\Delta_{m,th}^+}
\newcommand{\DeltaVlo}{\Delta_{m,th}^-}

\begin{document} 

\title{{\Large \textbf{Using published bid/ask curves to error dress spot electricity price forecasts}}}

\normalsize
\author{Gunnhildur H. Steinbakk, Alex Lenkoski\thanks{Corresponding Author.  This research was supported by The Norwegian Research Council 237718 through the  \emph{Big Insight} center for research-driven innovation.}\\  Ragnar Bang Huseby,
  Anders  L\o land\\ and Tor Arne \O ig\aa rd\\
\emph{Norwegian Computing Center}}
\vspace{2.5 in}
\maketitle

\begin{abstract}
  Accurate forecasts of electricity spot prices are essential to the daily operational and planning decisions made by power producers and distributors.  Typically, point forecasts of these quantities suffice, particularly in the Nord Pool market where the large quantity of hydro power leads to price stability.  However, when situations become irregular, deviations on the price scale can often be extreme and difficult to pinpoint precisely, which is a result of the highly varying marginal costs of generating facilities at the edges of the load curve.  In these situations it is useful to supplant a point forecast of price with a distributional forecast, in particular one whose tails are adaptive to the current production regime.  This work outlines a methodology for leveraging published bid/ask information from the Nord Pool market to construct such adaptive predictive distributions.  Our methodology is a non-standard application of the concept of error-dressing, which couples a feature driven error distribution in volume space with a non-linear transformation via the published bid/ask curves to obtain highly non-symmetric, adaptive price distributions.  Using data from the Nord Pool market, we show that our method outperforms more standard forms of distributional modeling.  We further show how such distributions can be used to render ``warning systems'' that issue reliable probabilities of prices exceeding various important thresholds.

\end{abstract}
\clearpage
\section{Introduction}
Forecasting electricity spot prices is a central concern for the operation of electricity producers and consumers. 
Typically the focus has been on deriving point forecasts, while probabilistic forecasts are more difficult and have been given less attention \citep{Weron2014, Nowotarski&Weron2017, Ziel&Steinert2018}. 
Precise and realistic forecasts of the electricity price distribution offer considerably more detail, which is beneficial to operators in the electricity sector. For example, deriving an optimal bidding strategy and generation plan requires understanding the full range of potential price outcomes. In this paper we discuss one method for constructing such price distributions which uses newly available data on bidding behavior.

Electricity is not a storable commodity, and the power system requires a constant balance between consumption and production. In addition, the demand for electricity depends on weather conditions and industrial production, which varies with the seasons, the weekdays and the holidays. Thus, the electricity market exhibits a unique price dynamic compared to other commodities, with seasonality at the daily, weekly and annual levels. In particular, unforeseen short-lived positive and negative price spikes are common occurrences in the electricity market. These features provide strong motivations for the use of distributional forecasts of electricity prices over point forecasts.

Rather than estimating the full distribution,
prediction of intervals (quantiles) have more commonly been used to study the variability in electricity prices \citep{Weron2014}. 
The prediction intervals have typically been estimated by bootstrapping \citep[e.g.,][]{Wan&etal2014, Khosravi&etal2013, Chen&etal2014},
or as quantiles of the error distribution, either from empirical prediction errors \citep[e.g.,][]{Misiorek&etal2006, Weron&Misiorek2008} or from some (semi)parametric model \citep[e.g.,][]{Panagiotelis2008, Weron&Misiorek2008, Zhao&etal2016,Dudek2016}. 
Other approaches for probabilistic electricity price forecasts are based on variants of GARCH models for describing conditional price densities, \cite[see e.g.,][]{Garcia-Martos&etal2011,Diongue&etal2009}.
Recently, methods have built on modifications and/or expansions of quantile regressions \citep[e.g., ][]{JonsonEtAl2014,Nowotarski&Werona2015,Maciejowska&etal2016,Gaillard&etal2016,Juban&etal2016}.
For a more thorough review of state of the art of probabilistic forecasts of electricity prices see \cite{Nowotarski&Weron2017}. 

Electricity spot prices are determined in an auction process on the power exchange. 
For a given delivery hour, the ask curve consists of all submitted offers to sell electricity at given volume and price.
Similarly, the bid curves are all the demand orders from different agents, each with its own quantity and price.
The hourly electricity spot price is then defined (modulo technical, system-level considerations) as the point where the bid and ask curves intersect.
The bid and ask curves have since year 2012 been publicly available at Nord Pool for the system as a whole, and the price is referred to as the system price. 

In this paper we combine
published bid/ask curves and point predictions to compute probabilistic forecasts for hourly day-ahead spot electricity prices. 
Modeling a time dependent bidding structure of supply and demand data at an electricity exchange directly is a difficult task \citep{Ziel&Steinert2016}. Instead our approach uses the error in volume predictions to form the basis for describing the variability, as the volume distribution is a more well-behaved distribution than that of the price distribution. Probabilistic prices and volumes are coupled by a forecast of bid/ask curves, which are themselves relatively stable in shape day-to-day.
The bid/ask curves contain information on the (aggregated) range in bidding and asking from market players and not just the points where supply and demand are matched. Hence, our approach is able to make more accurate forecasts of extreme spikes and sudden changes from one hour to the next.
Our methodology can be seen as a new form of ``error dressing'' \citep{roulston2003} by using the curves to translate residual behavior of market volume forecasts into price uncertainties.

The remainder of the paper is organized as follows. 
A description of background data from the Nord Pool market is given in section \ref{sec:data}, illustrated through data plots and simple empirical analysis.
This motivates the indirect probabilistic forecast model for electricity spot prices in section \ref{sec:methods}, the so-called bid/ask model.
Also, section \ref{sec:methods} describes two probabilistic benchmark models and a set-up for probabilistic forecast verification.
Results from the bid/ask model are given in section \ref{sec:results}, including comparison to the two benchmark models.
The final Section \ref{sec:conclusion} contains some concluding discussion.


\section{Background data\label{sec:data}}
The electricity market is made up of a large number of players that have differing levels of volume and marginal cost requirements. The market is settled via an auction process where the electricity price is determined from equilibrium between aggregated supply and demand.
In a day-ahead market, different buyers bid their desired volume and maximum price they would be willing to pay, delivered at a certain hour the next day.
All accumulated offers from these buyers form the bid curve, which then represents the relationship between price and volume for a given delivery hour for the buyers. 
Likewise, the ask curve contains all prices that the suppliers are willing to sell at a certain amount of electricity.
The electricity price is then calculated from the intersection of accumulated supply and demand. 
The bid/ask curves are published shortly after the spot market is settled in the Nordic power exchange Nord Pool.
Furthermore, agents submit their bids and offers for delivery of electricity each hour of the next day before noon. Thus, all prices in a day-ahead market are decided at the same time using the same available information.

\begin{figure}[!htbp]
\centering
\includegraphics[width=0.43\textwidth]{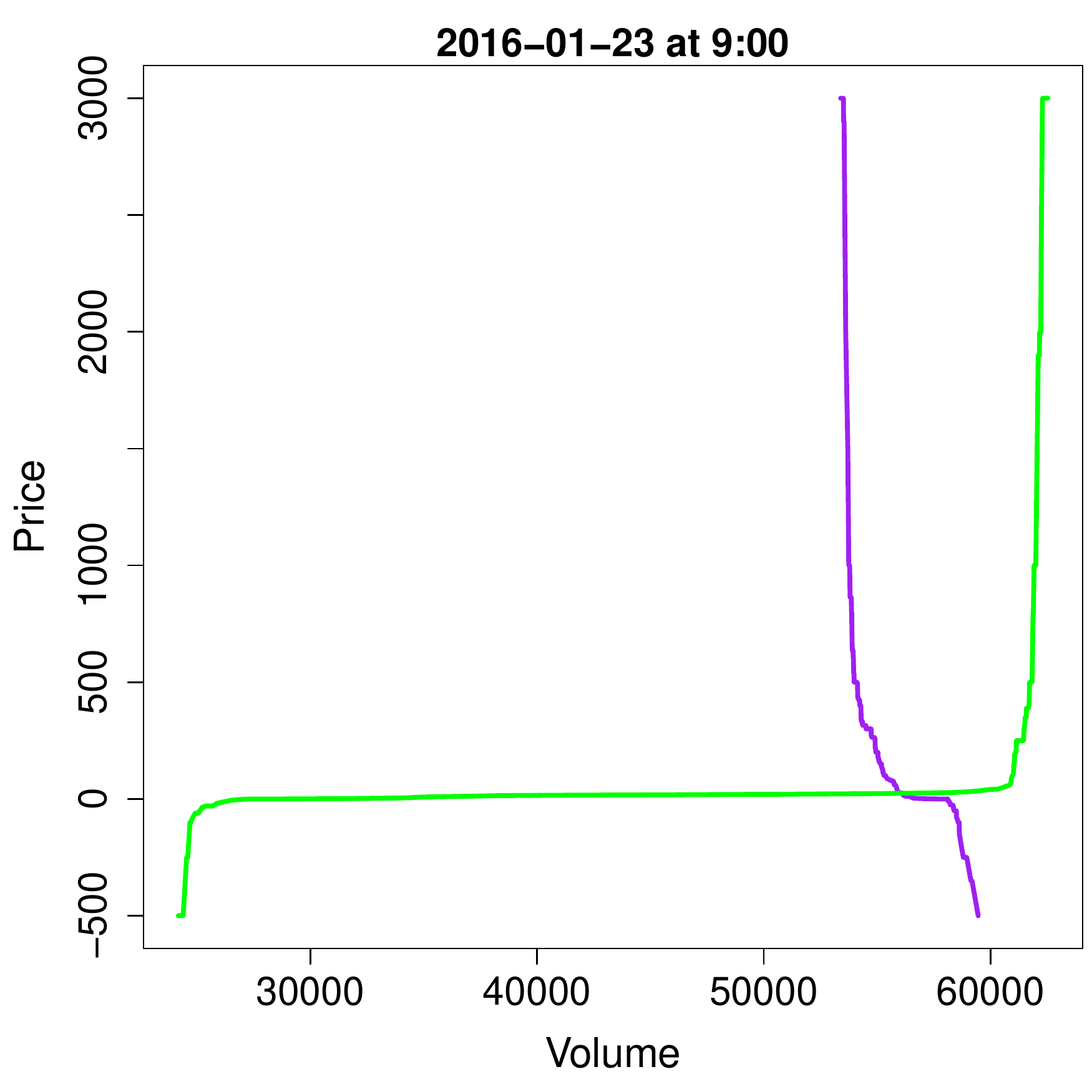}
\includegraphics[width=0.43\textwidth]{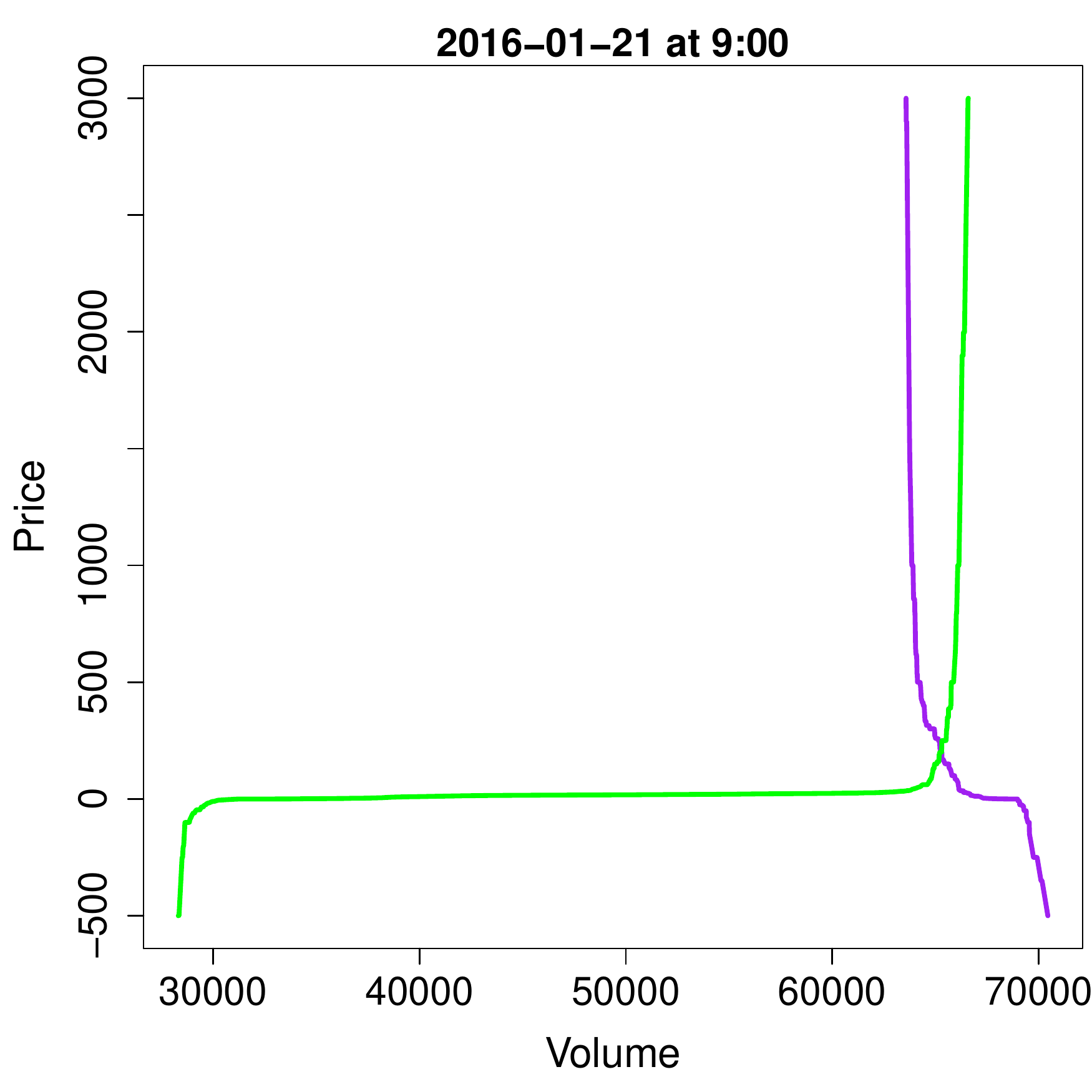}
\includegraphics[width=0.43\textwidth]{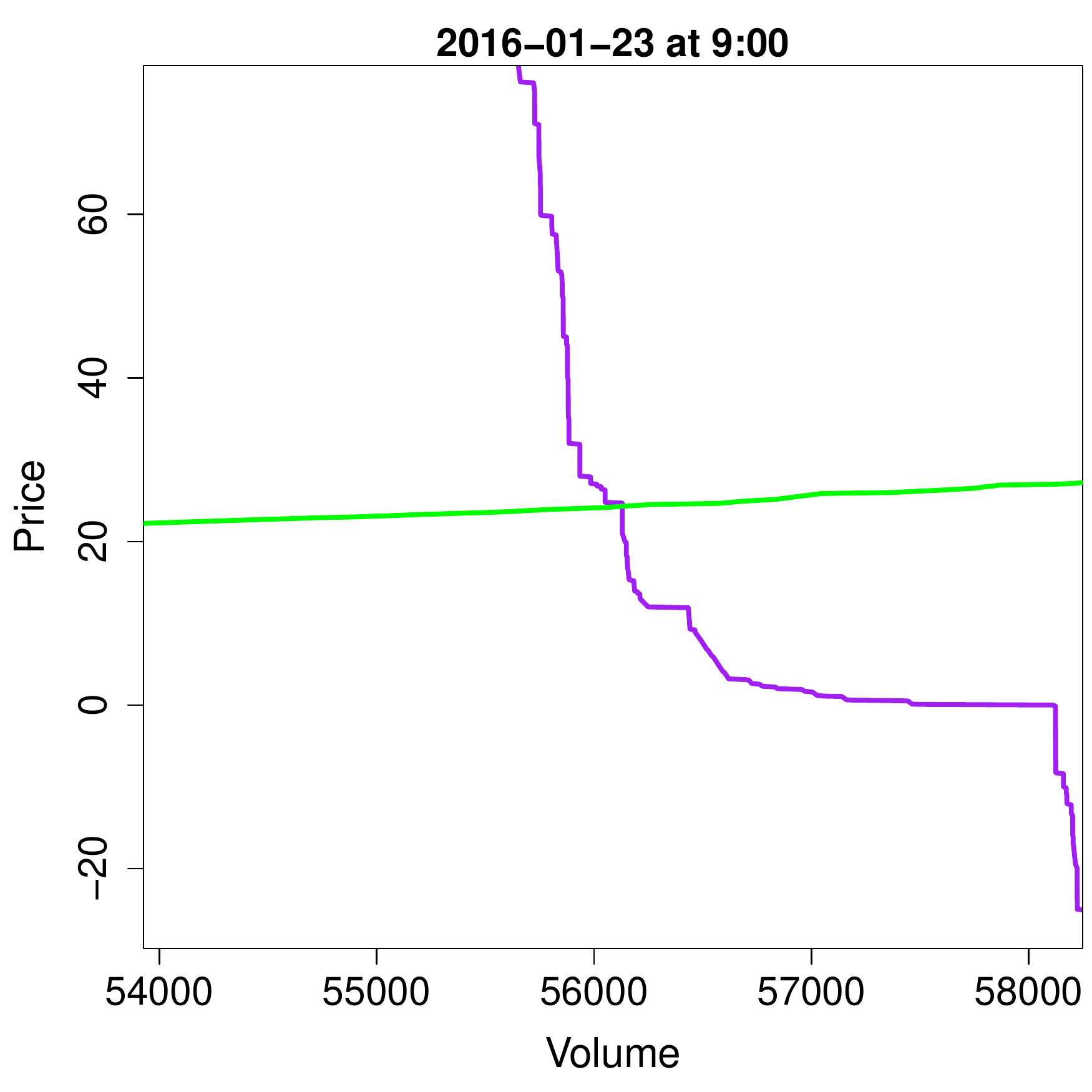}
\includegraphics[width=0.43\textwidth]{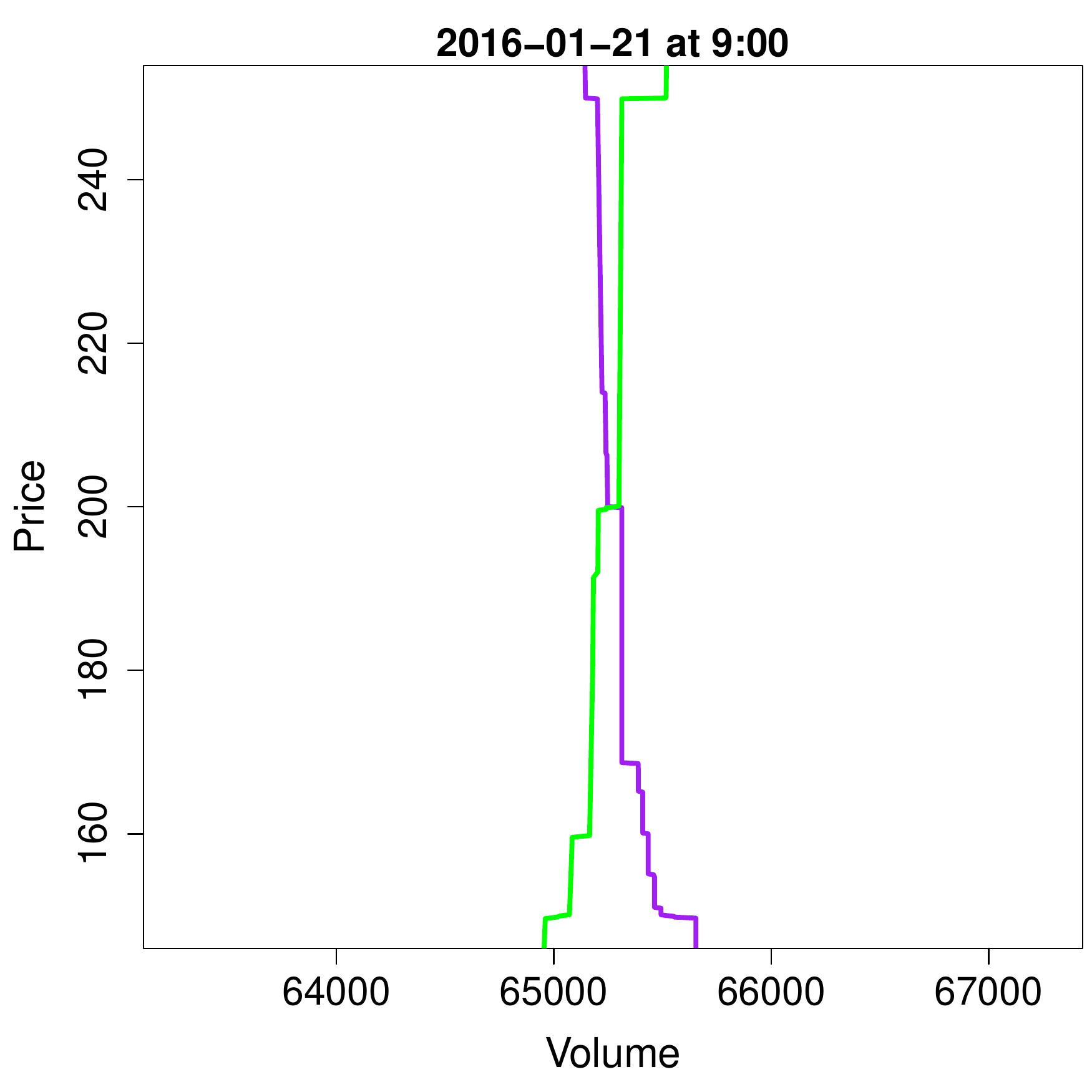}
\caption{Ask curve (green) and bid curve (purple) in a normal situation (left) and in an extreme situation (right). Top panel show the full range of bidding while the bottom panel shows area in the vicinity of intersection between the two curves.
}
\label{fig:bidask-curves}
\end{figure}

Nord Pool's major power source is hydro power, which is highly flexible, but has low marginal production costs. Roughly speaking, the hydro power producers can choose between producing today or storing the water.
In addition, there is a smaller number of units that can be started quickly but with high marginal cost.
The result is that the ask curve is mostly flat with a steep increase in the end of the curve.
The left panel of Figure \ref{fig:bidask-curves} shows a typical behavior of the bid/ask curves at Nord Pool. The slope of the ask curve is relatively flat around the point where the two curves intersect.  
Shifts in the bid curve will in those situations have minor effect on prices. 
In contrast, the right panel shows a rather extreme situation. The bid curve crosses the ask curve at a large volume where the slope of the ask curve is steep, due to the high marginal cost for producing this amount of electricity. The probability of high price spikes for equivalent shift in volume is much higher in such extreme situations as small changes in volume result in dramatic price changes.

The algorithm for coupling all submitted demands and supplies to derive the system area price is complex (see https://www.nordpoolspot.com/) and not fully disclosed to market participants in order to prevent gaming.
Furthermore, the final construction of these curves depends on a number of additional elements, such as block bids, cross-system flows and the curve starting point. All of these additional factors affect the absolute location of the curves, and considerable day-to-day variability is evident.
However, the relative shape of the curves is often similar from day to day.
Also, in practice, the two curves are step-functions, see Figure  \ref{fig:bidask-curves}, as not all prices are traded each day.

Throughout our exposition, we assume that a high-quality point forecast of electricity spot-price is available to the researcher and the goal is to ``error dress'' this point forecast, i.e. create a distributional forecast that is in some manner centered on the point forecast.  In our examples, we use a model developed by the authors in collaboration with Norsk Hydro ASA over the past two decades for point forecasts.  The specifics of this model are suppressed but are also not of particular relevance.  The error dressing methodology discussed here would function equally well with any other high-quality forecast.  We note, however, that simplistic ``straw man'' point forecasts (e.g. a persistence forecast) introduce too much error and therefore it is necessary to discuss the performance of our proposed methodology in the context of a high-quality point forecast.

\begin{figure}[!htbp]
\centering
\includegraphics[width=0.9\textwidth]{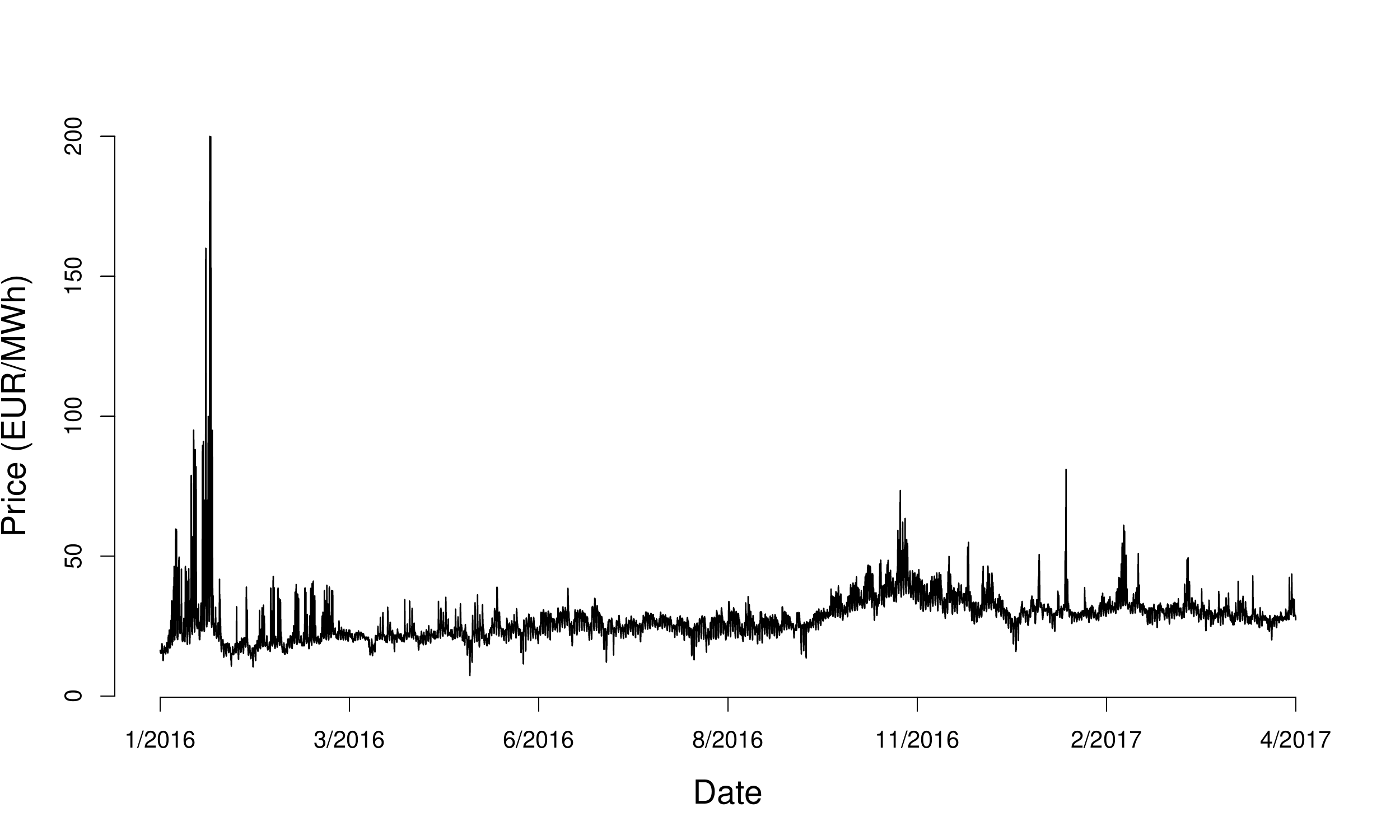}
\includegraphics[width=0.9\textwidth]{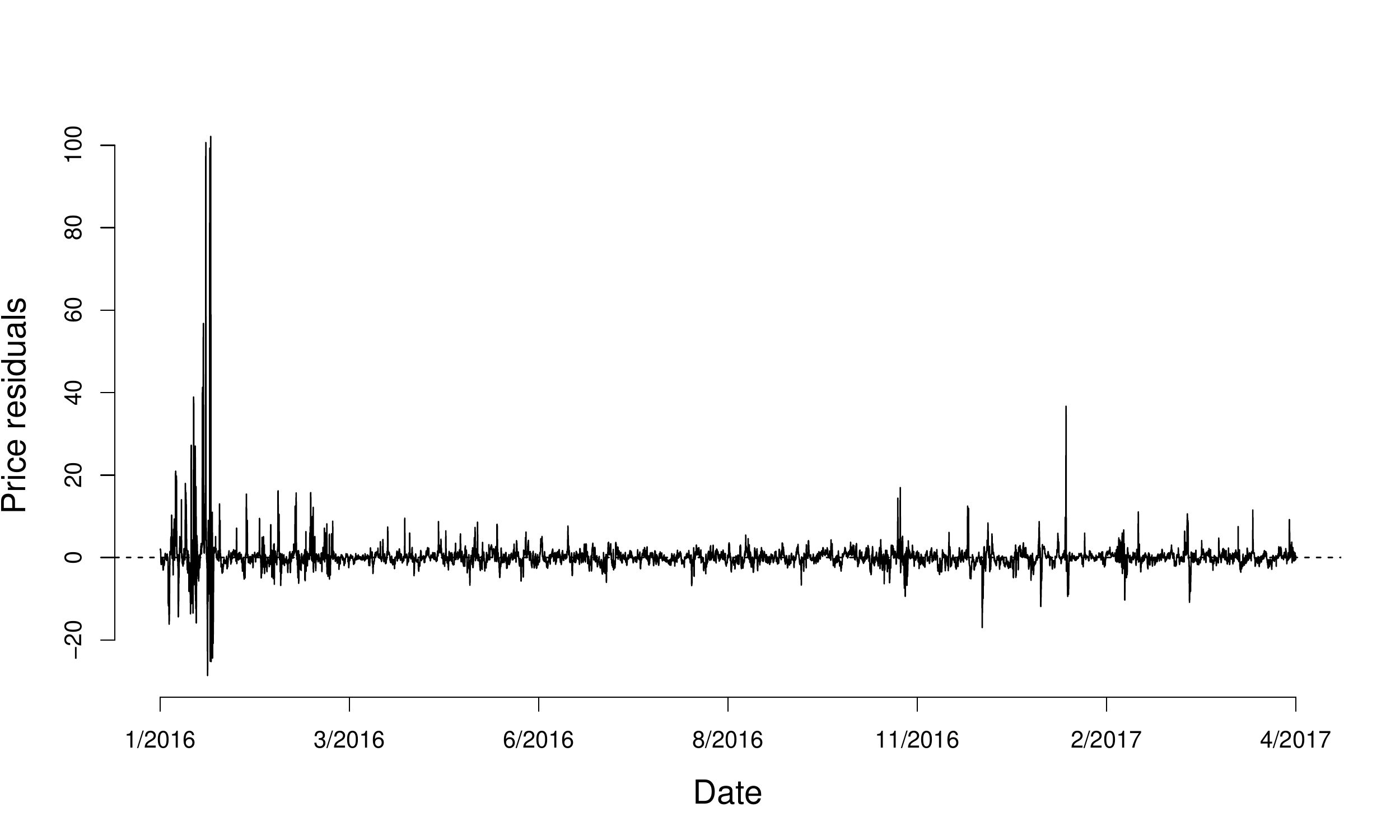}
\caption{Electricity system spot prices (upper panel) and residuals of spot price predictions, using data from Nord Pool from January 2016 to April 2017.}
\label{fig:el-price}
\end{figure}

 The upper panel in Figure \ref{fig:el-price} shows a time series plot of the hourly system prices for electricity in Nord Pool from January 2016 to April 2017, illustrating several features of electricity price market, such as spikes and seasonal variations.  The large  price spikes from beginning of year 2016 is in particular prominent. The lower panel in Figure \ref{fig:el-price} shows hourly price residuals, computed as the differences between the observed price and our corresponding day-ahead price prediction for each hour.  The variations in mean and volatility are less severe for these price residuals than for the time series of prices, but the spikes are still visible. 

\begin{figure}[!htbp]
\centering
\includegraphics[width=0.9\textwidth]{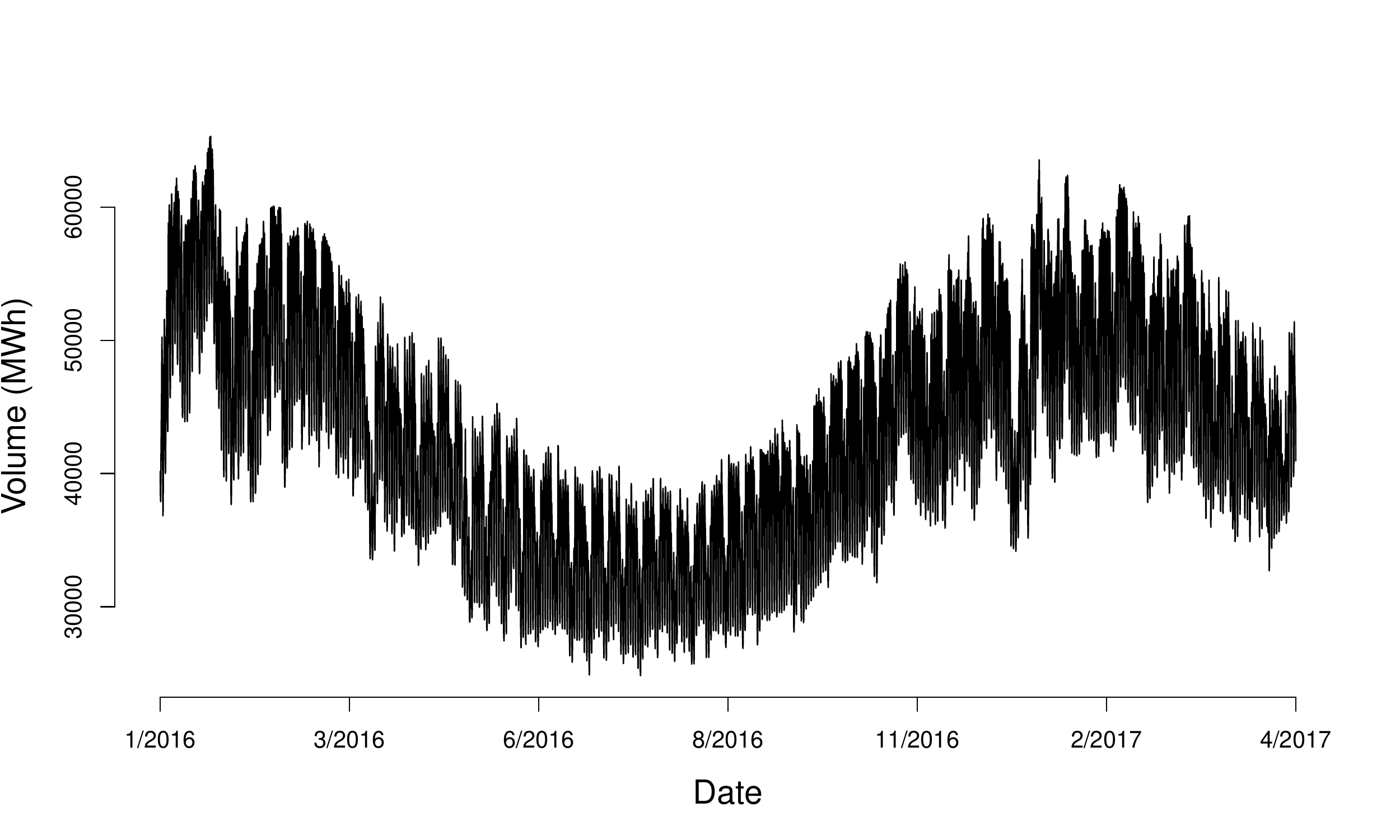}
\includegraphics[width=0.9\textwidth]{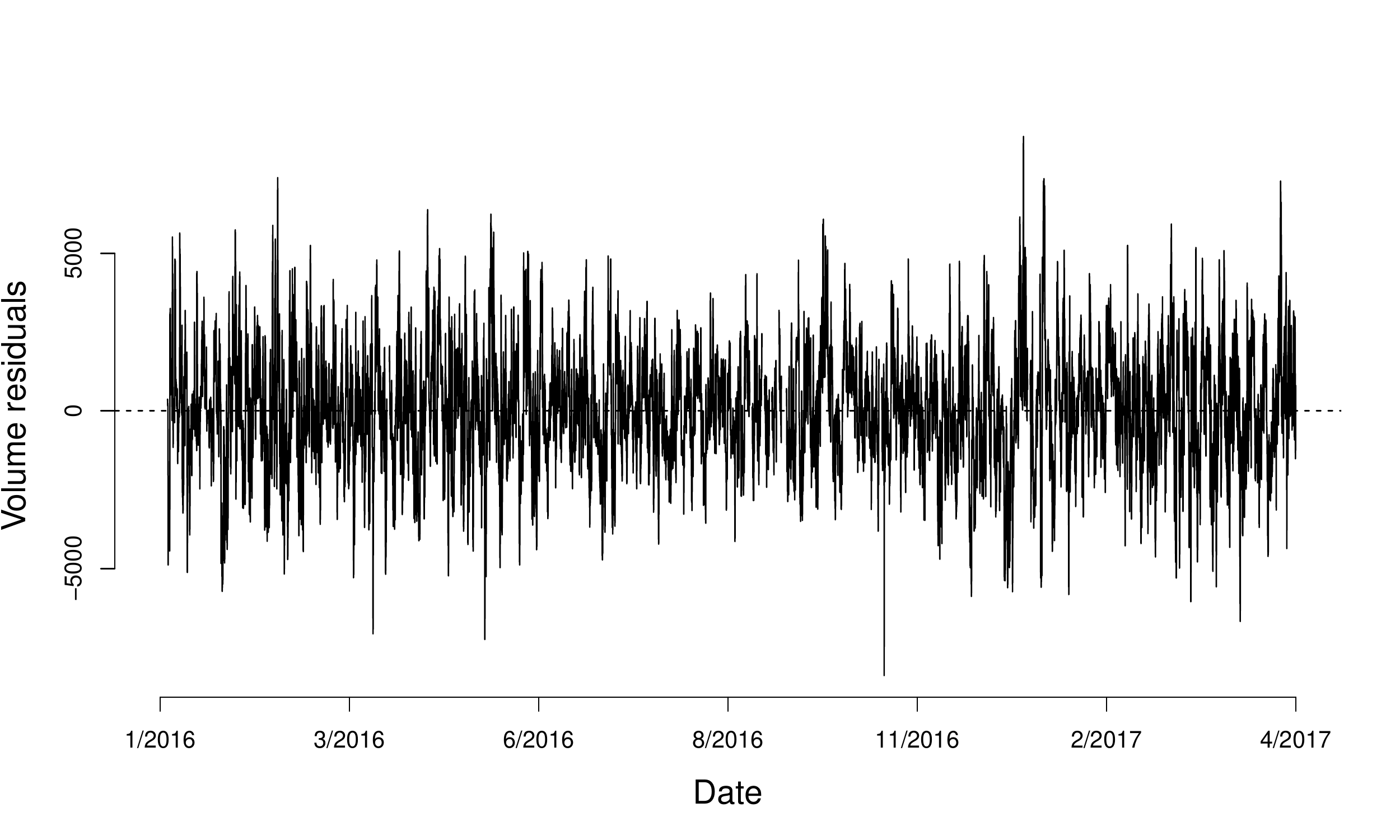}
\caption{Electricity volume and residual of volume predictions, using data from Nord Pool from January 2016 to April 2017.}
\label{fig:vol}
\end{figure}

\begin{figure}[!htbp]
\centering
\includegraphics[page=9,width=0.45\textwidth]{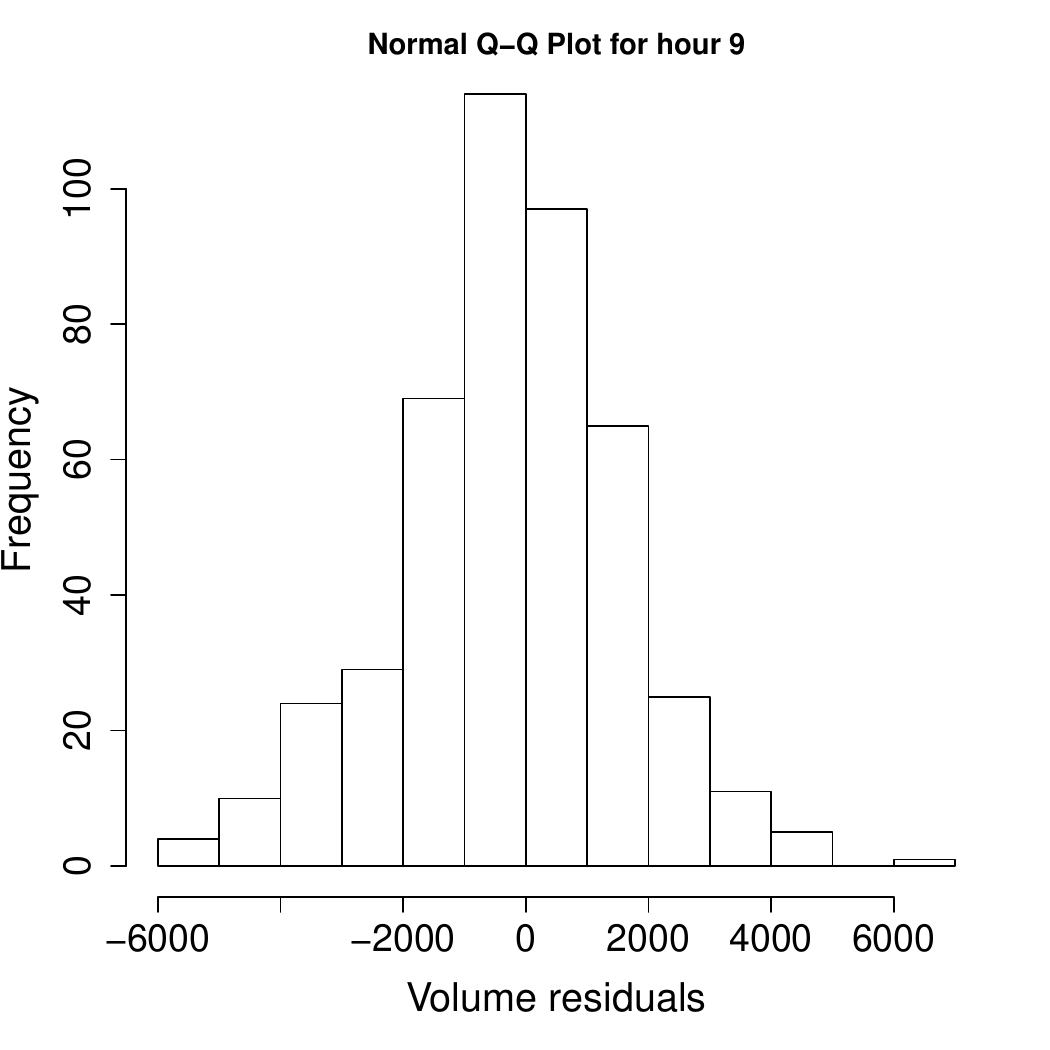}
\includegraphics[page=10,width=0.45\textwidth]{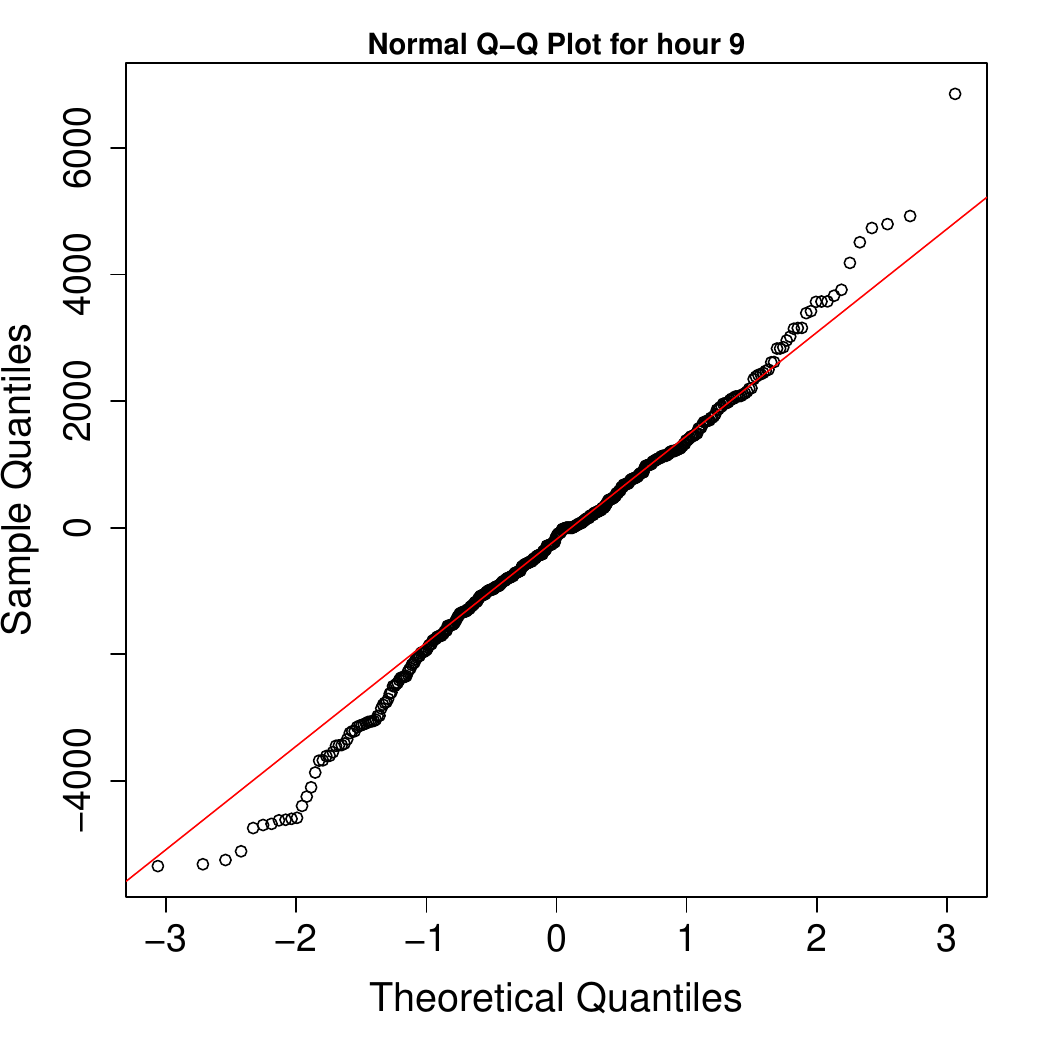}
\caption{The empirical distribution and qq-plot of volume residuals at hour 9, using data from Nord Pool from January 2016 to April 2017.}
\label{fig:volhist}
\end{figure}

The bid and ask curves describe the relation between price and volume. For a given price estimate, we may use the ask (or the bid) curve to compute the corresponding volume estimate. 
Hence, instead of defining a separate model, the volume is modeled indirectly by using the curves to transform price estimates to volume estimates, as described in section \ref{sec:methods}.
Unlike the electricity prices, the dynamic of electricity volume in the upper row of Figure \ref{fig:vol} has a structured pattern with systematic daily, weekly and seasonal variations and little evidence of sudden spikes.
The distributionally stable behavior of the volume residuals is even more clear in Figure \ref{fig:volhist}, which shows the empirical distribution and a Q-Q plot of volume residuals at hour 9. 
A rather simple Gaussian distribution is thus appropriate for modeling the volume residuals at a specific hours.
The clear auto-correlation in the lower row of Figure \ref{fig:volhist}, due to intra-day dependencies, will be considerably smaller when considering only one hour at a time in the Gaussian model for volume distribution.  

\begin{figure}[!htbp]
\centering
\includegraphics[width=0.8\textwidth,page=1]{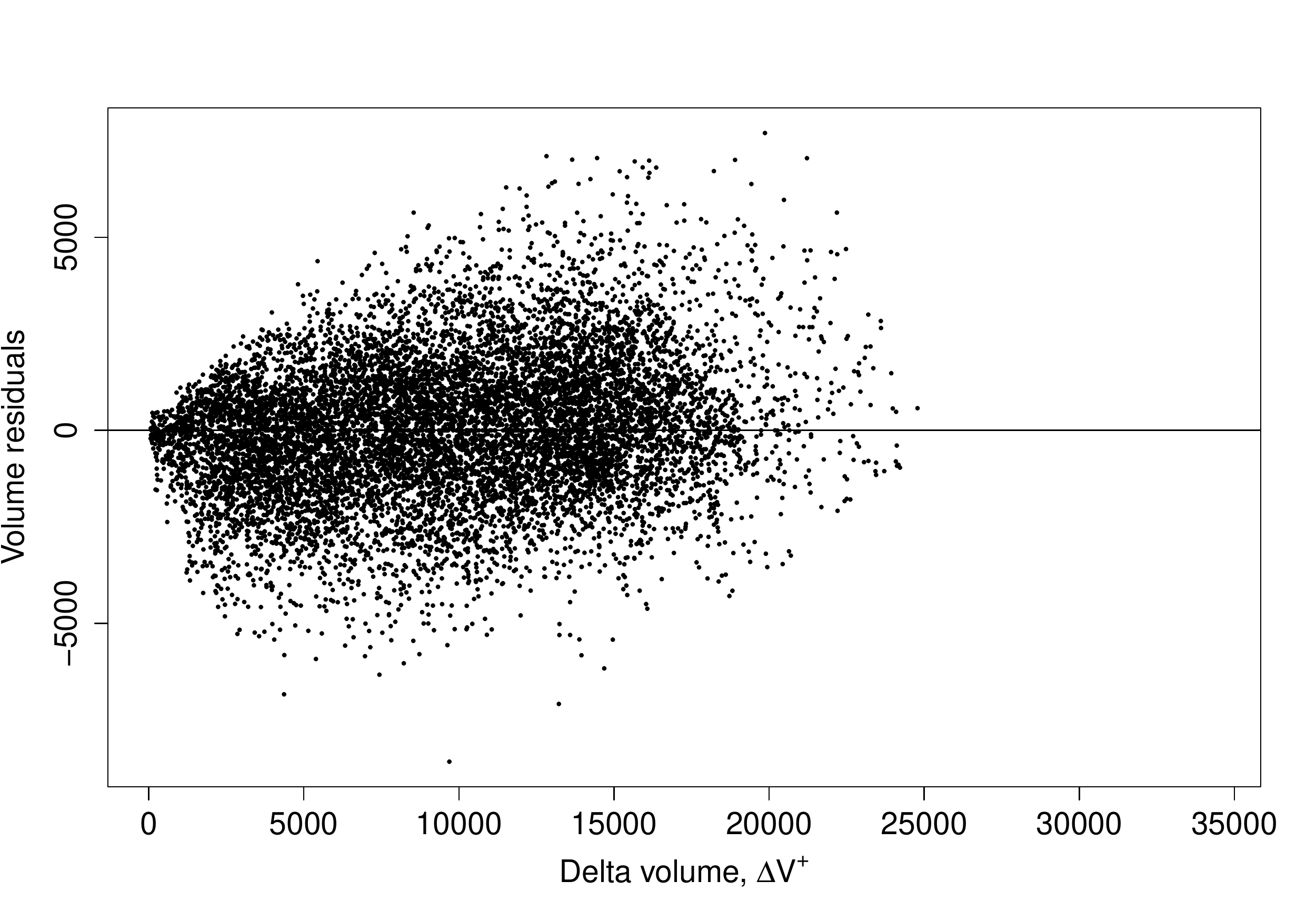}
\includegraphics[width=0.8\textwidth,page=2]{./volumeErrorPlots.pdf}
\caption{
The spread of volume error against the absolute change in the volume, where the changes in volume correspond to a certain change in the price estimates on the bid curves. 
Upper row:  The price estimates are increased by 50\euro, i.e., $\Delta_{50,{th}}^+$ defined in \eqref{eq:deltavol_+} is displayed on the x-axis. Lower row: The price estimates are decreased by 50\euro, i.e, $\Delta_{50,{th}}^-$ defined in \eqref{eq:deltavol_-} is displayed on the x-axis.}
\label{fig:deltavolume}
\end{figure}

While volume residuals are distributionally well-behaved, they are not homoscedastic. In particular, as the forecasted price rises, the spread of volume error diminishes.  We capture this behavior by constructing a descriptive feature of the curve.  In particular, for a given ask curve, we compute the corresponding change in volume as we increase the price estimate by a certain amount. These changes in volume will be smaller if the price estimate is close to the steep end of the ask curve. The upper panel of Figure \ref{fig:deltavolume} shows the absolute changes in volume as we increase the price estimates by 50\euro, illustrating that spread of the (positive) volume error are smaller for small changes in volume. A decrease of 50\euro~from the price estimates, does not, on the other hand, show such a pattern in the change of volume except for a few cases, see the lower panel of Figure \ref{fig:deltavolume}.


\begin{figure}[!htbp]
\centering
\includegraphics[width=0.7\textwidth]{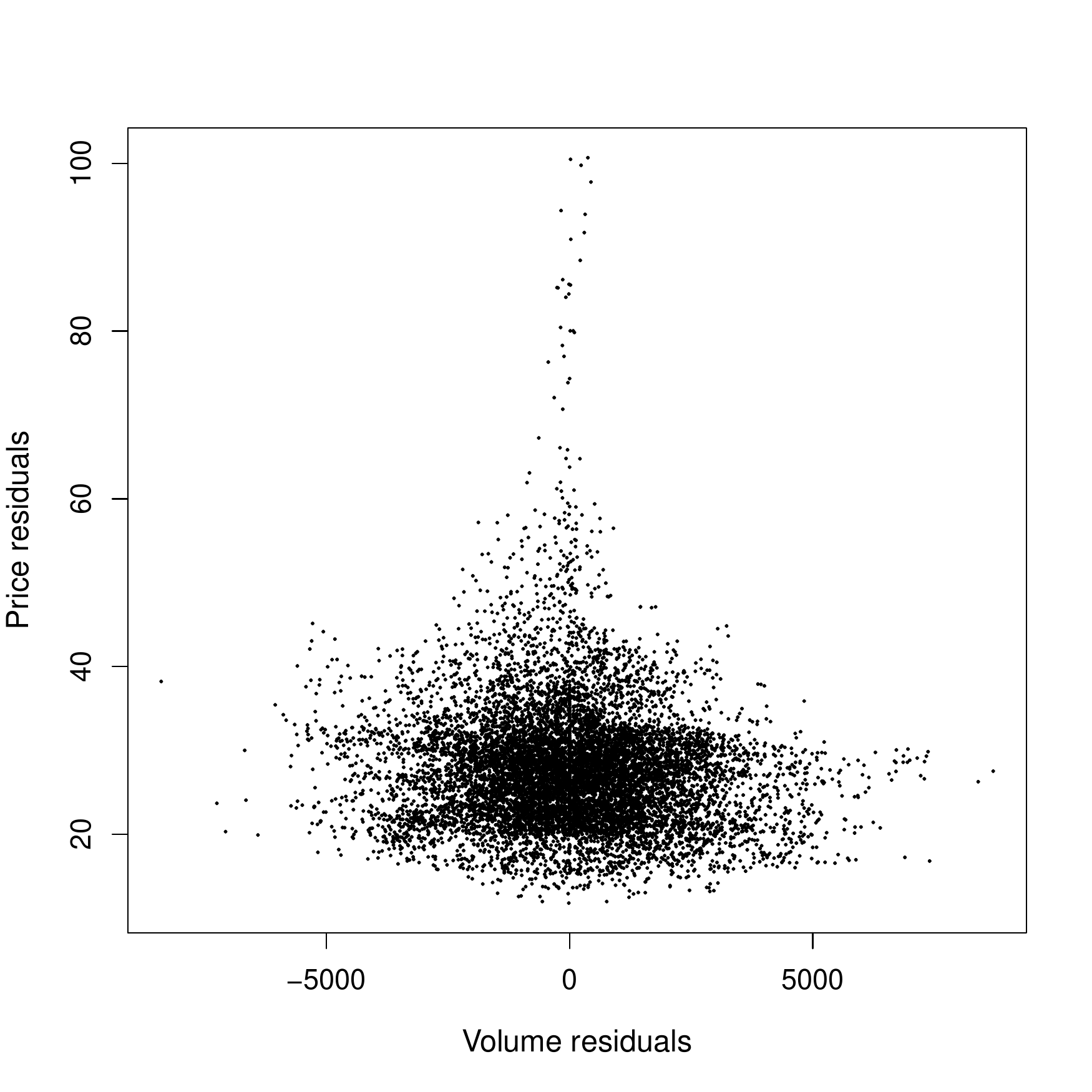}
\caption{The volume residuals against the estimated prices, based on data from Nord Pool from January 2016 to April 2017.}
\label{fig:scatterPriceVolError}
\end{figure}

\newpage

\section{Methodology\label{sec:methods}}

\subsection{Indirect model of electricity prices using bid/ask curves and volume uncertainty}
Let $p_{th}$ and $v_{th}$ be the electricity price and the volume, respectively, at day $t$ and hour $h$. 
Further, let $b_{th} : \mathbb{R}_{+} \rightarrow  [-500,3000]$ be a decreasing function describing the bid (or buy) curves.
Similarly, define the ask (or sell) curve as the increasing function $s_{th} : \mathbb{R}_{+} \rightarrow  [-500,3000]$.
The electricity price and volume are then defined as the points where the bid and ask curves intersect. Thus,
\[
v_{th} = \arg \min_{v \in \mathbb{R}_{+} } |b_{th}(v) - s_{th}(v)|
\quad \mbox{ and } \quad 
p_{th} = s_{th}(v_{th}).
\]
The functions above define how the prices and volumes are related through the bid/ask curves.

Let $\mathcal{D}_t = \{ p_{th}, v_{th}, b_{th}, s_{th} \}$, for hours  $h=1,\dots,24$, be the set of prices, volumes and bid/ask curves at day $t$,
and let $\mathcal{X}_t$ be all necessary auxiliary information (such as weather predictions) at time $t$ for hours $h=1,\dots,24$.
Furthermore, denote by $\mathcal{T}_t$ all the data given up to time $t$, this is 
$\mathcal{T}_t =  \{ \{\mathcal{D}_s\}_{s<t}, \{\mathcal{X}_s\}_{s<t}\}$.

Denote by $\hat p_{th}$ the price forecast for day $t$ and hour $h$, generated by a point forecasting model the day before $t$,
given the data $\mathcal{T}_t$. The volume predictions may then be indirectly computed as $\hat v_{th} = s_{th}^{-1}(\hat p_{th})$, where $s_{th}^{-1}(\cdot)$ is the inverse sell curve. This gives us historic volume residuals $e_{sh} = \hat v_{sh} - v_{sh}$ for days $s<t$ and for hours $h=1,\dots,24$.

Our approach for computing the distribution $F_{th}$ for the hourly prices $p_{th} \sim F_{th}$, given data $\mathcal{T}_t$ contains of two steps.
First, we compute the distribution $G_ {th}$ of the volume error $\epsilon_{th}$. 
The distribution of the hourly electricity prices is then defined as a functional transformation of the volumes $v_{th} = \hat v_{th} + \epsilon_{th}$, where the transformation is based on the bid/ask curves. Hence,
\begin{eqnarray*}
  \epsilon^{(r)}_{th} &\sim& G_ {th},\\
  p_{th}^{(r)} &=& s_{th}(\hat v_{th} + \epsilon^{(r)}_{th}).
\end{eqnarray*} 
Here, $G_{th}$ will be estimated from historic residual $e_{sh} = \hat v_{sh} - v_{sh}$ for $s<t$ and all hours  $h=1,\dots,24$.
The volume residuals are relatively symmetric and well-behaved (see Figure \ref{fig:volhist}) and thus $G_{th}$ may typically be a rather simple distribution, such as Gaussian.
However, due to the heteroscedasticity displayed in Figure \ref{fig:deltavolume}, the variance of the distribution is set to react to where the forecasted price $\hat{p}_{th}$ lands on the bid curve.
 
In particular, define 
\begin{equation}\label{eq:deltavol_+}
\DeltaVup = |s_{th}^{-1}(\hat p_{th} + m) - s_{th}^{-1}(\hat p_{th})|, 
\end{equation}
which is the absolute change in volume as we increase the price forecast $\hat p_{th}$ by $m$. Similarly, we define the absolute change in volume as we decrease the price forecast by $m$, 
\begin{equation}\label{eq:deltavol_-}
\DeltaVlo = |s_{th}^{-1}(\hat p_{th} - m) - s_{th}^{-1}(\hat p_{th})|. 
\end{equation}

\begin{figure}[!htbp]
\centering
\includegraphics[width=0.47\textwidth,page=1]{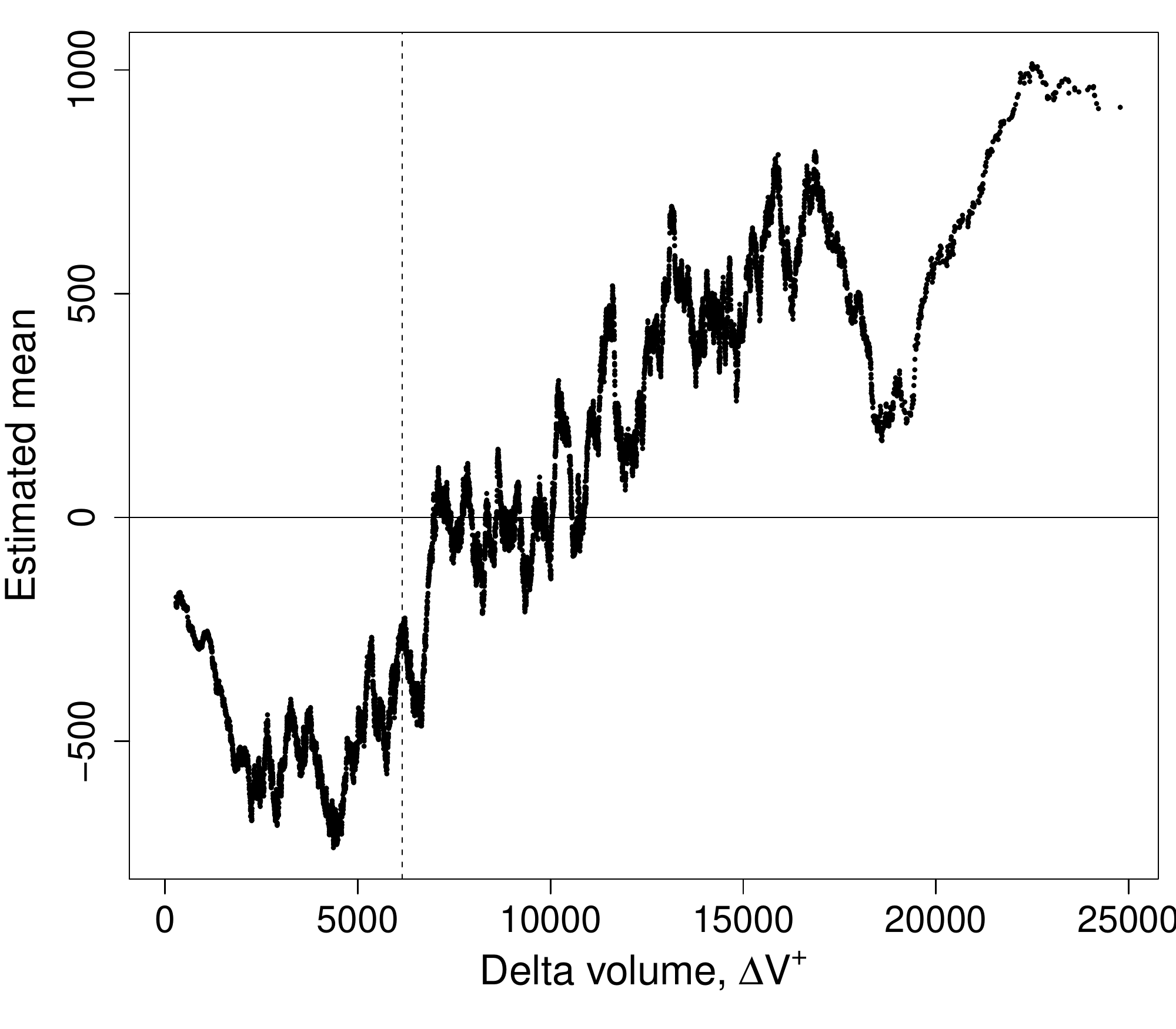}
\includegraphics[width=0.47\textwidth,page=2]{./estimateMeanStdVolumError.pdf}
\caption{Estimated mean and standard deviation of the volume residuals, computed based on the 500 nearest neighbors with similar values of delta volume $\DeltaVup$.
The dashed vertical line is the threshold $\Delta_0=6150$ used in the analysis in section \ref{sec:results}.}
\label{fig:deltaErrorMeanStdev}
\end{figure}

If we sort all historic volume errors according to the size of $\DeltaVup$ for a given $m$, say $m=50$, 
we may estimate the sample mean and standard deviation based on all volume errors with similar values of $\DeltaVup$. In Figure \ref{fig:deltaErrorMeanStdev} we used a moving window of 500 neighbors, which illustrates that this moving sample variance of the volume distribution decreases as $\DeltaVup$ goes to zero, while the sample mean is always negative if the volume changes $\DeltaVup$ are small.
As expected such a pattern does not exists for $\DeltaVlo$ (not shown here),  as Nord Pool's role as an energy exporting market ensures price collapses occur rarely.
Thus, the volume error distribution is assumed to have different mean and standard deviation if $\DeltaVup$ is below a threshold $\Delta_0$
than if $\DeltaVup$ is greater than $\Delta_0$.

Denote by $\gamma^2_{th}$ the variance of the volume distribution for small volume changes $\DeltaVup \leq \Delta_0$ and
$\tau^2_{th}$ as the variance for volume errors with volume changes larger than the threshold $\Delta_0$. Then the variance of the volume error is given by
\begin{equation}\label{eq:sigma2}
\sigma_{th}^2 = \mathbbm{1}_{\{ \DeltaVup > \Delta_0\}}\tau_{th}^2 + \mathbbm{1}_{\{\DeltaVup \leq \Delta_0\}}\gamma_{th}^2.
\end{equation}
Similarly, we define the expectation of the volume error, $\mu_{th}$, as a sum of two expectations depending on if the price estimates  satisfy $\DeltaVup \leq \Delta_0$ or $\DeltaVup > \Delta_0$.

\subsection{Benchmark models}
The performance of our model is compared to that of two benchmark models. The distribution of these two benchmark models is estimated based on the empirical price residuals $r_{th} = \hat{p}_{th} - p_{th}$ for day $t$ and hour $h$. 

The first benchmark model is given by $p_{th} = \hat{p}_{th} + \sigma_{th} z_{th}$, with $z_{th}$ as i.i.d. zero mean Gaussian variable with standard deviation 1. The parameter $\sigma_{th}$ is estimated from historical data, that is 
\[
\hat{\sigma}_{th} = \sqrt{ \frac{1}{|\mathcal{N}_t|-1} \sum_{\substack{s<t \\ 1\leq h \leq 24}} (p_{s,h} - \hat p_{s,h})^2},
\]
where $|\mathcal{N}_t|$ is the number of observations up to day $t$. Here, the electricity prices are error dressed directly, while our proposed approach uses electricity volume residuals and then translate volume forecasts to price uncertainty by the bid/ask curves. 

The other benchmark model is given by $p_{th} = \hat{p}_{th} + \mathbf{r}_{th}^{*}$, where $\mathbf{r}_{th}^{*} = \{r_{sh}\}$ is the set of all historical price residuals $r_{s,h}$ for $h = 1,\dots,24$ and $s<t$.
The estimated spread of the distribution of these two benchmarks are thus constant up to day $t$. 
The first mentioned benchmark is referred to as the ``Gaussian benchmark model'',  while the other as the ``Empirical benchmark model''.

\subsection{Forecast validation and verification\label{sec:validation}}
We aim to assess and compare reliability (calibration) and sharpness of the models \citep{Gneiting&Raftery2007}. A calibrated model gives a precise estimate of outcome uncertainty while sharpness addresses the concentration of the forecast distribution.

A popular reliability assessment tool is the probability integral transform (PIT), often displayed graphically by histograms. 
 If observations follow the estimated predictive distribution $\hat F$, the PIT will be uniformly distributed \citep{Dawid1984}
\[
\hat F \sim U([(0,1)]).
\]
Continuous ranked probability score (CRPS) measures the difference between the forecast and observed CDFs \citep{Gneiting&Raftery2007},
\[
\mathrm{CRPS} = \int_{-\infty}^{+\infty} (\hat F(x) -  \mathbbm{1}_{\{p < x\}})^2 dx,
\]
where $p$ is the realized observed price.
To assess of the reliability in the tail of the predictive distribution, we apply the quantile score, 
see e.g. \citep{Friederichs&Hense2007}, \citep{Gneiting&Raftery2007} and references therein, given as
\[
s_Q(\hat F,p|\tau) = (p - \hat F^{-1}(\tau))(\tau - \mathbbm{1}_{\{ p \leq \hat F^{-1}(\tau)\}}),
\]
for a specific probability level $\tau \in (0,1)$.

The predictive power is evaluated by a cross validation in time,  
so that the training data set consist of only observations that occurred prior to the 24 hours that forms the validation data set. 
This scheme is repeated for every day during the observation period.
The forecast accuracy is then computed by averaging over all validation data sets.

\section{Results\label{sec:results}} 
This section shows results from the aforementioned bid/ask model to predict probabilistic forecasts of hourly day-ahead spot electricity prices.
The results from the bid/ask model are then compared to that of the Gaussian and the Empirical benchmark models.

For predicting probabilistic spot prices for a specific day $t$ and hour $h$ based on the bid/ask model, we used the most recently available bid curve for hour $h$, which usually means the previous day's curve at hour $h$.
The predictive volume error distribution is assumed to follow a Gaussian distribution, with a mean and standard deviation as described in Section \ref{sec:methods}.
The standard deviations with price estimates satisfying $\DeltaVup > \Delta_0$, i.e. $\tau_{th}$ in \eqref{eq:sigma2}, are estimated by computing the sample standard deviations based on all historic
volume errors $e_{sh}$ 
for hour $h$ the last 120 days. 
The standard deviations $\gamma_{th}$s in \eqref{eq:sigma2} are estimated by a moving sample standard deviation based on all historic volume errors where $\Delta_{m,sh}^+ \leq \Delta_0$ for $s<t$ and \textit{all} hours $h=1,\dots,24$. 
In particular, we sort the set of $\{\Delta_{m,sh}^+\}$, $s\leq t$ and $h=1,\dots 24$, in an ascending order. For a given $\DeltaVup$, we find the window of the 100 neighbors of the sorted $\Delta_{m,sh}^+$s and compute the sample standard deviations of the volume errors that corresponds to these 100 neighbors.
Similar approaches were used for computing the sample mean $\mu$.

Several other values for the number of neighbors were tested, but a window of 100 neighbors gave the best prediction performance (in the sense of lowest value of CRPS and quantile scores). 
By inspection of Figure \ref{fig:deltaErrorMeanStdev}, we used the threshold $\Delta_0 = 6150$. 
In addition, we set $m=50$ and used the previous day's ask curves to estimate the volume changes $\DeltaVup$.
Using other values of $\Delta_0$ and $m$ did not improve the predictive performance. 

\begin{figure}[!htbp]
\centering
\includegraphics[width=0.95\textwidth,page=1]{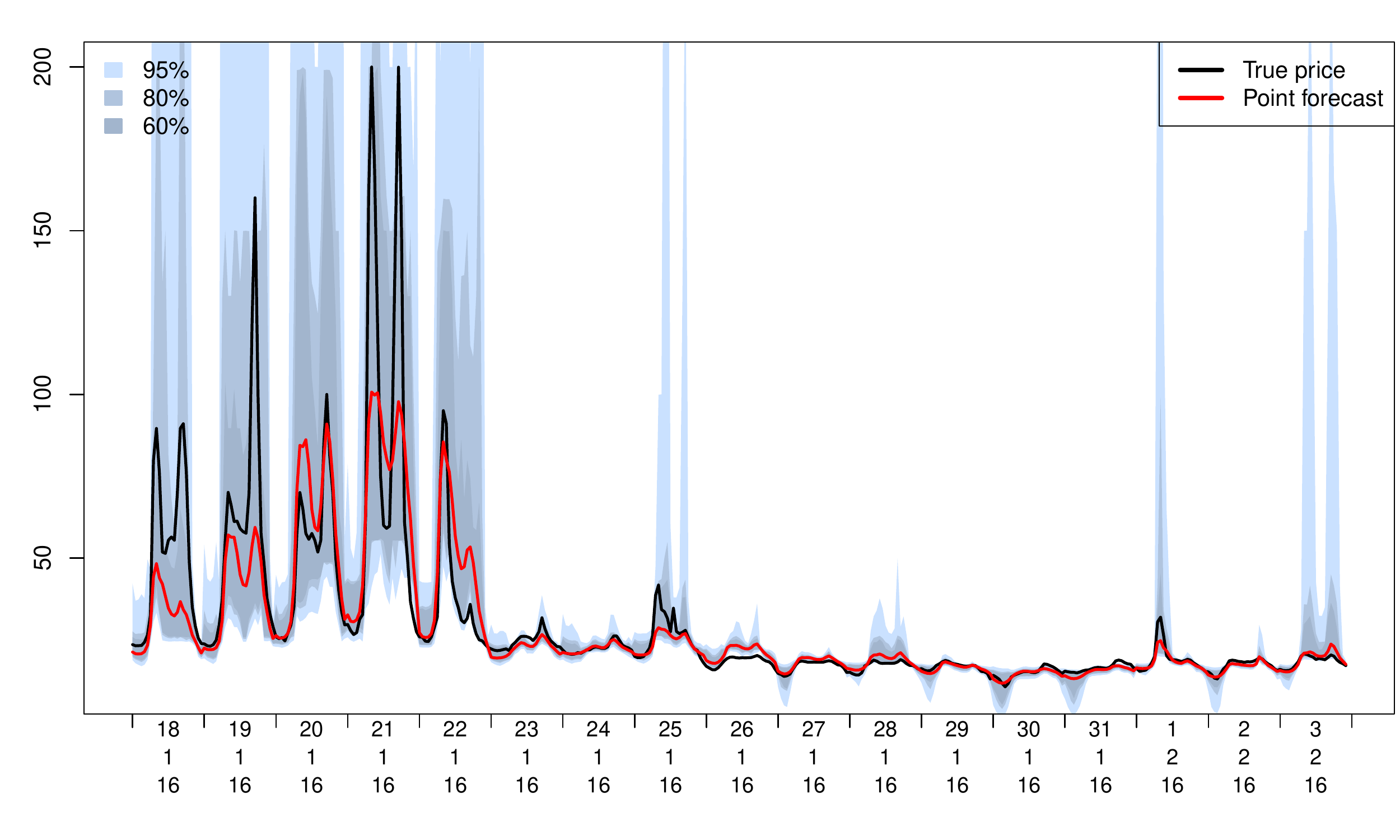}
\includegraphics[width=0.95\textwidth,page=12]{./priceBandsBidAsk.pdf}
\caption{Probabilistic forecasts of electricity spot prices based on the bid/ask model for 17 days in January 2016 (upper row) and in July 2016 (lower row), displaying the 60\%, 80\% and 95\% quantiles of the probabilistic distribution. The black line is electricity spot price $p_{th}$ and the red line is electricity spot forecasts $\hat p_{th}$.}
\label{fig:timeSeriesPred}
\end{figure}

Figure \ref{fig:timeSeriesPred} shows probabilistic forecasts from the bid/ask model for two different time periods. 
The extreme high prices at 21 January 2016 in the upper row are captured by the uncertainty bands, although the price point forecast was by no means near such high prices. Also, the plot in lower row illustrates that the dipping prices are within the 95 \% uncertainty band. 

\begin{figure}[!htbp]
\centering
\includegraphics[width=0.45\textwidth,page=1]{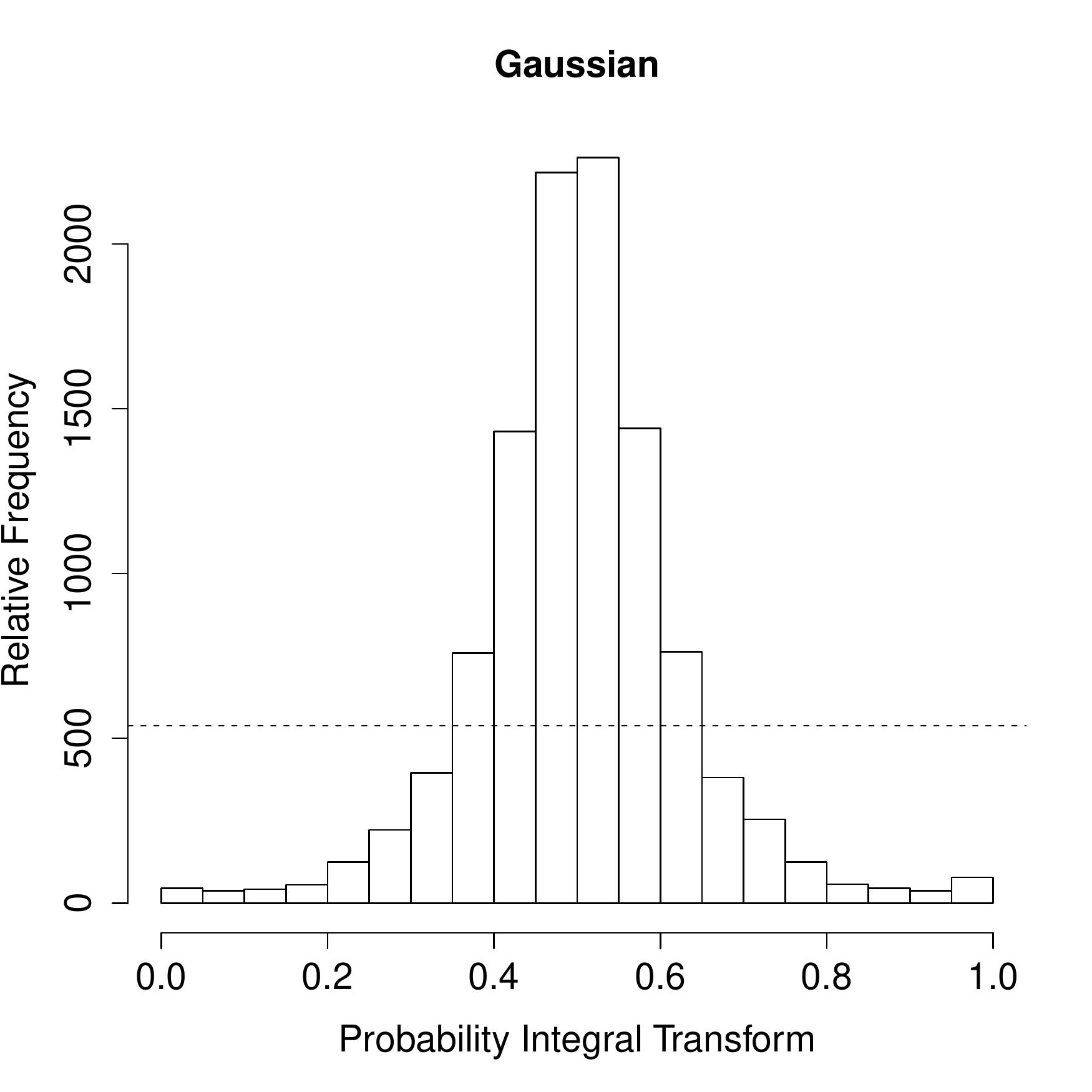}
\includegraphics[width=0.45\textwidth,page=2]{./PIT-2.pdf}
\includegraphics[width=0.45\textwidth,page=3]{./PIT-2.pdf}
\caption{Probability integral transform (PIT) histograms for the benchmark models and the bid/ask model.}
\label{fig:PIT}
\end{figure}

We validate the models following Section \ref{sec:validation} and start by comparing histograms of all PIT values for the bid/ask model and the two benchmark models (Gaussian and Empirical) in Figure \ref{fig:PIT}. 
The Gaussian benchmark model's average over all 24-hours prediction intervals have far to high variance, as the histogram forms a clear hump in the middle of its plot.  The Empirical model have on average also under confident predictions, but much less severe than than the Gaussian model.

In contrast, Figure \ref{fig:PIT} shows that the PIT values of 
the bid/ask model's average over all 24-hours prediction intervals are quite well calibrated. The leftmost column indicates that the variance in the left tail distribution is a bit to small, while the upper side of the distribution has a good fit.
We are only taking into account the absolute volume changes in the upper right bid curve when estimating the standard deviation in the volume error.
Thus, we might in some cases underestimate the standard deviation in the left tail of the price distribution as the absolute volume changes are different for the lower part than for the upper part of the bid curve.

Table \ref{tab:scores} displays the average CRPS and quantile scores for the cross-validation predictions, comparing the bid/ask model to that of the two benchmark models. The bid/ask model performs better than the benchmark models. 
In order to investigate the significance of the scores for the bid/ask model and the empirical model in Table \ref{tab:scores}, 
we perform a permutation test following the procedure in \cite{Molleretal2013}.

This permutation test proceeds by assuming that the two populations of CRPS scores come from the same distribution under the null-hypothesis. For each hour we randomly assign one score to the Empirical model and the other to the bid/ask model. 
We make 10 000 samples of synthetic data sets by randomly pairwise re-assigning the scores from the two models each hour
and compute the difference in mean between the two re-assigned populations.
The 0.025 and 0.975 quantile of these differences in means are -0.01515  and 0.01551, while the observed population is 0.04644, which is far away from any values in the permutation data set. 
The magnitude of the difference in CRPS in Table \ref{tab:scores} is not large, however. This shows that the results could not have come from sampling variability only. 
 
\begin{figure}[!htbp]
\centering
\includegraphics[width=0.45\textwidth]{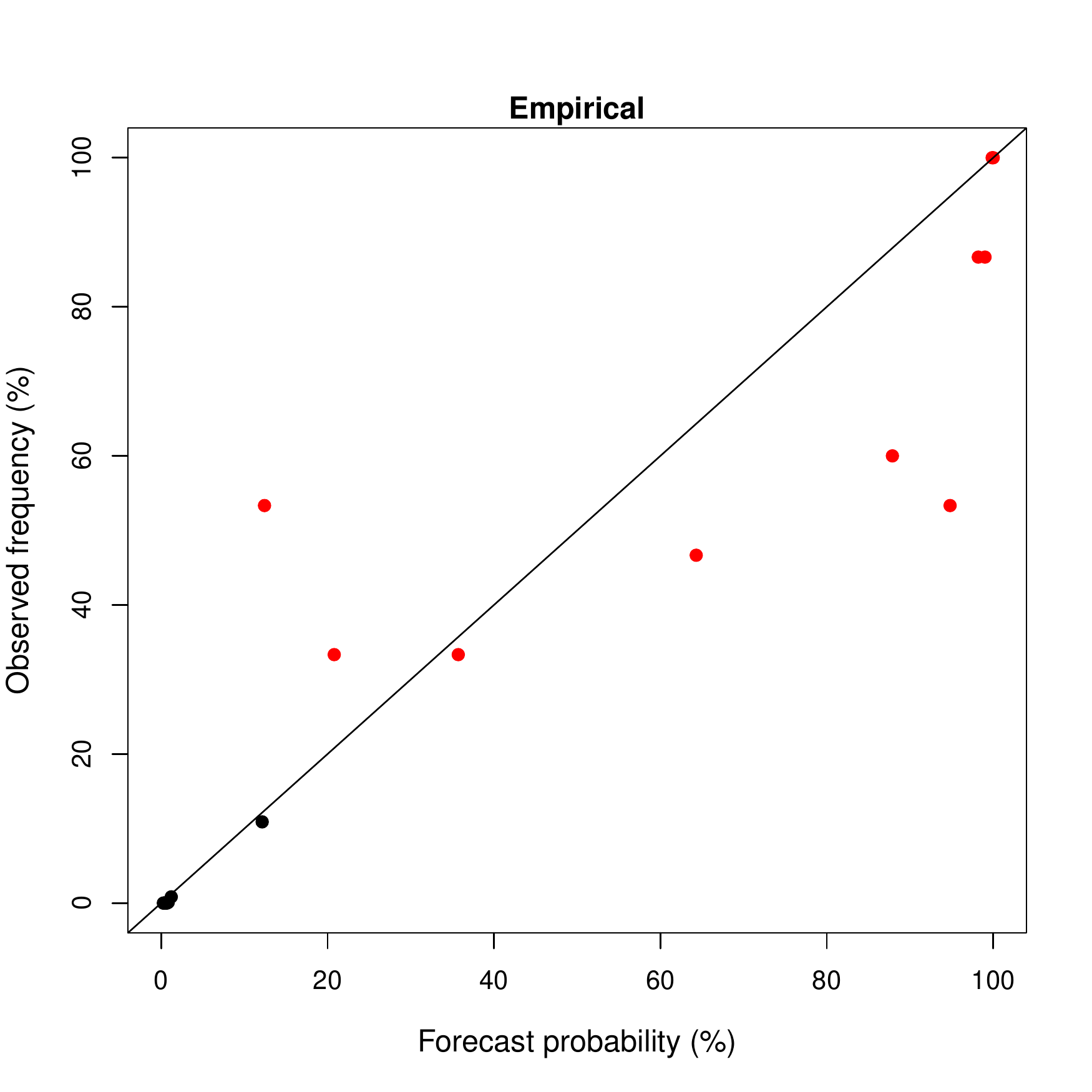}
\includegraphics[width=0.45\textwidth]{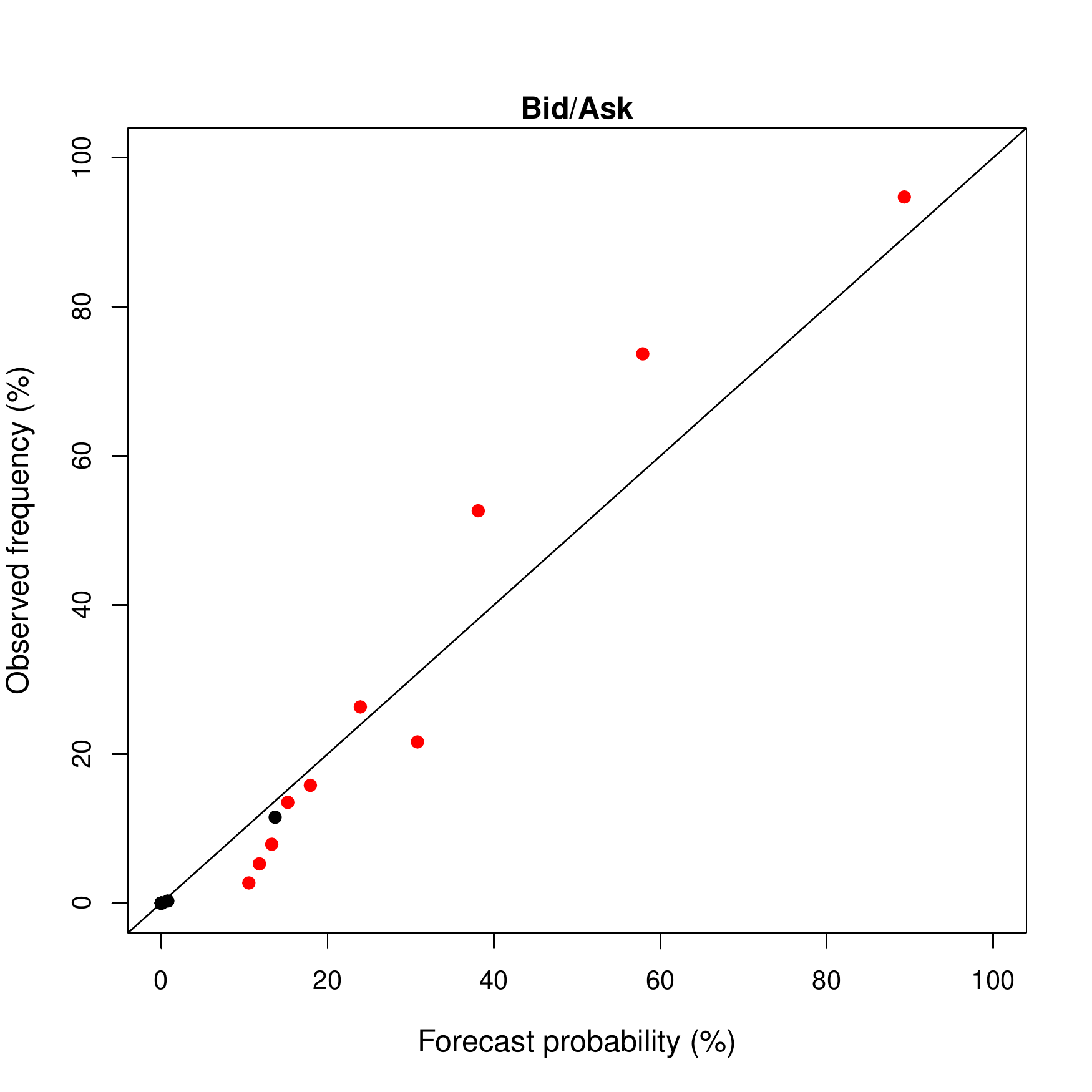}
\caption{Reliability plots for the probability of exceeding 50\euro, i.e Pr$\{p>50\mbox{\euro}\}$ for prices $p$, for the Empirical benchmark model and the bid/ask model. The red dots show the reliability for the subset of estimated Pr$\{p>50\mbox{\euro}\}$ greater than 0.1.}
\label{fig:calibration}
\end{figure}

Conducting a similar permutation test for the quantile score at probability level $\tau = 0.1$ gave a similar clear conclusion,
but the difference between the quantiles scores at $\tau = 0.9$ is not significant. 
However, the bid/ask model is better to capture the probability of the extreme prices. 
Figure \ref{fig:calibration} displays reliability plot of the forecast probabability of exceeding prices of 50\euro,
where the range of the forecasted exceedance probabilities were divided into 10 equally sized bins. 
Prices exceeding 50\euro~is a quite extreme situation and happens in 1.14\% during all hours in our data set. 
The performance of the bid/ask model and the Empirical benchmark model seems to be equally successful when including all outcomes of the exceedance probabilities (see the black dots in Figure \ref{fig:calibration}).
In order to zoom into the tail of this highly skew distribution of the exceedance probabilites, only the forecasted exceedance probabilities greater than 10\% are considered in the plotted red dots.
Here, we clearly see that the probabilistic forecasts of extreme situations from the bid/ask model is considerably better calibrated than the Empirical benchmark model.
Precise and reliable exceedance probabilities can be important for a warning system to forecast hours or periods with potentially extreme prices.
 
There is still some work in progress for estimating proper probabilistic forecast in all situations. The bid/ask model can sometimes for regular prices give to high probabilistic forecasts variances if the price predictions are at the steep end of the bid curve, all though the CRPS in Table \ref{tab:scores} and the PIT plot in Figure  \ref{fig:PIT} show that the overall fit is good.

\begin{table}[!htbp]
  \begin{center}
    \begin{tabular}{l c c c}
\hline\hline
 & Gaussian & Empirical & Bid/Ask\\ 
\hline
$\overline{\mathrm{crps}}$ & 1.60 & 1.05 & 1.00\\ 
$\overline{s_Q}(\tau=0.1)$ & 0.72 & 0.36 & 0.27\\ 
$\overline{s_Q}(\tau=0.9)$ & 0.78 & 0.45 & 0.43\\ 
\hline\hline 
\end{tabular}

    \label{tab:scores}
  \end{center}
\caption{Mean scores of CRPS and quantile scores for the 90 \% and 10 \% for the two benchmark models (Gaussian and Empirical) and the bid/ask model.}
\end{table}

\section{Conclusion\label{sec:conclusion}}
In this paper, we presented a method for computing probabilistic forecasts for hourly day-ahead spot electricity prices.  Our methodology addressed the heavy tailed nature of electricity price distributions by incorporating a new source of data, namely published bid/ask curves. As we have detailed, the curves themselves are unsuited to direct modeling, and are best used as a system for error dressing an existing point forecast of electricity prices.  We then focused modeling on the easier task of forecasting the distribution of volume errors that result from a given point forecast.  By deriving a feature that described the proximity a given price forecast is to the ``kink'' in the ask curve we were able to handle residual error heteroscedasticity in a parsimonious manner.  Our error dressing system outperforms other distributional forecasting methods, especially when modeling tail events.

The methodology outlined here is intended to be incorporated into a much larger market-modeling system and a number of practical additions have not been addressed.  Chief amongst these is cross-border trade of electricity.  In particular, Nord Pool trades electricity with other systems operating in the Netherlands, Germany, Poland, among others.  Each of these markets have similar price resolution systems and are themselves connected to even more European markets further South and East.  Many (but not all) of these systems supply their own bid/ask data.  An important additional feature to develop is a coupled error dressing system that models these markets jointly and thereby adjusts prices accounting for the potential of cross-border flow.

In addition, one element that has yet to be investigated is the forecast of the bid and ask curves themselves.  In practice, we have found that while the curve as a whole can shift greatly, the relative structure of the curve is reasonably consistent day-to-day.  This in part motivated our use of the bid/ask curves in the context of error dressing (which recenters the system around the point forecast).  However, a more involved probabilistic model of curve evolution may yield refined price distributions.

Finally, the spot market settles all hours in the following day simultaneously.  At present we focused on marginal error dressing methodologies which modeled each hour's volume error individually.  In practice, these residuals are clearly likely to be correlated.  Extending the model into a multivariate forecasting system should prove straightforward.

\section*{Acknowledgements}
The author's gratefully acknowledge the helpful support and advice of Ellen Paaske and Stefan Erath of Norsk Hydro ASA.  

\bibliographystyle{plain}
\bibliography{ref}

\end{document}